\documentclass[12pt]{article}
\usepackage{amsmath}
\usepackage{latexsym}
\usepackage{amsfonts}
\usepackage[normalem]{ulem}
\usepackage{array}
\usepackage{amssymb}
\usepackage{graphicx}
\usepackage[backend=biber,
style=numeric,
sorting=none,
isbn=false,
doi=false,
url=false,
]{biblatex}

\usepackage{subfig}
\usepackage{wrapfig}
\usepackage{wasysym}
\usepackage{enumitem}
\usepackage{adjustbox}
\usepackage{ragged2e}
\usepackage[svgnames,table]{xcolor}
\usepackage{tikz}
\usepackage{longtable}
\usepackage{changepage}
\usepackage{setspace}
\usepackage{hhline}
\usepackage{multicol}
\usepackage{tabto}
\usepackage{float}
\usepackage{multirow}
\usepackage{makecell}
\usepackage{fancyhdr}
\usepackage[toc,page]{appendix}
\usepackage[hidelinks]{hyperref}
\usetikzlibrary{shapes.symbols,shapes.geometric,shadows,arrows.meta}
\tikzset{>={Latex[width=1.5mm,length=2mm]}}
\usepackage{flowchart}\usepackage[paperheight=11.69in,paperwidth=8.27in,left=0.79in,right=0.79in,top=0.79in,bottom=0.79in,headheight=1in]{geometry}
\usepackage[utf8]{inputenc}
\usepackage[T1]{fontenc}
\TabPositions{0.5in,1.0in,1.5in,2.0in,2.5in,3.0in,3.5in,4.0in,4.5in,5.0in,5.5in,6.0in,6.5in,}

\setlistdepth{9}
\renewlist{enumerate}{enumerate}{9}
		\setlist[enumerate,1]{label=\arabic*)}
		\setlist[enumerate,2]{label=\alph*)}
		\setlist[enumerate,3]{label=(\roman*)}
		\setlist[enumerate,4]{label=(\arabic*)}
		\setlist[enumerate,5]{label=(\Alph*)}
		\setlist[enumerate,6]{label=(\Roman*)}
		\setlist[enumerate,7]{label=\arabic*}
		\setlist[enumerate,8]{label=\alph*}
		\setlist[enumerate,9]{label=\roman*}

\renewlist{itemize}{itemize}{9}
		\setlist[itemize]{label=$\cdot$}
		\setlist[itemize,1]{label=\textbullet}
		\setlist[itemize,2]{label=$\circ$}
		\setlist[itemize,3]{label=$\ast$}
		\setlist[itemize,4]{label=$\dagger$}
		\setlist[itemize,5]{label=$\triangleright$}
		\setlist[itemize,6]{label=$\bigstar$}
		\setlist[itemize,7]{label=$\blacklozenge$}
		\setlist[itemize,8]{label=$\prime$}

\begin{document}
{\fontsize{17pt}{20.4pt}\selectfont \textbf{\textit{Straddling between determinism and randomness:}}\par}\par

{\fontsize{16pt}{19.2pt}\selectfont \textit{Chaos theory vis-à-vis Leibniz.}\par}\par

{\fontsize{14pt}{16.8pt}\selectfont \textit{Vedad Famourzadeh\footnote{\  Freelance researcher, M.Sc. in Mechanical Engineering from Sharif University of Technology (vedad.famourzadeh@gmail.com) }, Maysam Sefidkhosh\footnote{ Assistant Professor of Shahid Beheshti University, Faculty of Theology and Religious studies,Visiting Professor of Philosophy of Science at Sharif University of Technology (m\_sefidkhosh@sbu.ac.ir) }}\par}\par

\vspace{\baselineskip}\textbf{Abstract: }The problematic relationship between science and philosophy has, since the beginning of modernity, divided the world into two separate domains- nature and human. Some of today’s schools of philosophy and epistemological inquiry have maintained a radical separation to the point where they refuse to maintain any commonality within the two. We argue that such a dichotomy will not only destroy the idea of the unity of knowledge from a theoretical perspective but, it will destroy a unified understanding of reality. This article is a critique of now mainstream belief of such dichotomy through re-reading of Leibniz’s idea on the unity of knowledge. The distinctive feature of Leibniz’s approach, which was specifically laid down in his correspondence with Clarke, is that his way of reasoning is philosophical as well as physical. In Leibniz’s view what guarantees the soundness of philosophical principles cannot defy the laws of physics. We shall also demonstrate that Chaos Theory is consistent with the affinities of Leibniz’s approach. The main focus of this article is on re-evaluating views on causal determinism and randomness as well as free will. The final claim of this article is that the natural world and human world follow the same causal deterministic laws, and that causal laws are however nonlinear. Knowledge of our reality should be rigorously sought in both the domains of humanities and natural sciences, but this causal determinism should distance itself from a simple linear causal model. What is thought of as random events or probabilistic events in the physical world, as well as in the human world, are not ontologically run by chance. The complexity and vastness of reality are such that our knowledge of reality always suffers a lack, and it is this lack of knowledge that is usually being called randomness.\par

\vspace{\baselineskip}\textbf{\textit{I. Introduction}}\par

\vspace{\baselineskip}Our question is how causal and deterministic laws govern our world and how accidentals can still be an outcome of a rational view of the world. The answer to this question will also define the relation of man to nature. The rational view results from a distancing from the world which separates humans from nature. This separation led to two domains, of nature and of human, and deterministic laws were only assigned to the domain of nature and even understanding random accidentals in this domain has been problematic. We argue that one can simultaneously embrace both determinism and random accidentals within a rational view of the world only if this rational view is overly complex and if an epistemic attribute is assigned to random accidentals and not an ontological value. A simple rational mind finds these two elements in a contradiction, and therefore can only accommodate one of these elements. The simple rationale mind also renders causality as simply linear and asks: Is the world governed by a linear causal chain of laws, or is the nature of the world accidental? This question has marked the history of philosophy beginning with some pre-Socratic philosophers and up until Leibniz, and the question persists today.\par

\vspace{\baselineskip}What is it that gives this world its order and structure? This world is believed to be governed by determinate, definite, rational laws that leave no room for irregularities. One reaches a dilemma:\ whether one has to constantly change and update one’s  account of the causal chain by explaining that our information is not enough to account for an explanation of the observed phenomenon and that there should be some unforeseen physical causes to such phenomena, or one could resort to an esoteric account of a world of mystery and postulate an intervening God regulating all orders and structures in our world, and therefore conceding to a mythological account of the world. But, explaining the world through a simple linear causal chain has at times produced shortcomings and leaves some inexplicable excesses of the system that seem disordered and unfit for a perfect causal account. \par

\vspace{\baselineskip}
Today modern physics rejects simple answers to classical physical as well as metaphysical way of reasoning. We believe instead of trying to overturn causality or rational organisation of things altogether, scepticism should aim at the very \textit{duality} of causal determinism and accidental happenings. In this article, we try to find a new approach inspired by Leibniz's philosophy which simultaneously avoids such dualism and contemporary relativism. We find this approach necessary for understanding the problematic relation between man and nature. This approach emerges employing the ideas of Leibniz. Leibniz can be interpreted as a philosophy of strong determinism while his deterministic view was highly complex. The elegance and uniqueness of his philosophy are in the interconnection of the world of natural physics and his metaphysical philosophy. As such, his philosophy is a turning point and an opening for both the philosophical and scientific way of thinking. We embrace both determinism and randomness by granting randomness to an epistemological domain, i.e. in the eye of the observer, while still adhering to strong determinism within the ontological domain. Before proceeding further, it is required to establish definitions of some key terms \par

\vspace{\baselineskip}
\begin{justify}
\textit{Determinism}: Determinism and causality are the ontological categories of a system. Laplace has given a clear concise classical definition of determinism. We find Laplace’s definition is more akin to Leibniz’s philosophy. Laplace writes:
\end{justify}\par

\vspace{\baselineskip}\begin{adjustwidth}{0.5in}{0.0in}
\begin{justify}
Given for one instant an intelligence which could comprehend all forces by which nature are animated and the respective situation of the beings who compose it – an intelligence sufficiently vast to submit these data to analysis - it would embrace in the same formula the movements of the greatest bodies of the universe and those of the lightest atom; for if, nothing would be uncertain and the future, as the past, would be present before its eyes$ \ldots $ 
\end{justify}\par

\end{adjustwidth}

\begin{adjustwidth}{0.5in}{0.0in}
\begin{justify}
The curve described by [the motion of] a molecule of air or of vapor is following a rule as certainly as the orbits of the planets: the only difference between the two is due to our ignorance. Probability is related, in part to this ignorance, in part to our knowledge. Bricmont (1995), (Marqui 2013)
\end{justify}\par

\end{adjustwidth}

\vspace{\baselineskip}
\textit{Probability}: For Laplace, probabilities are a way of rationally inferring the state of the system when we have not incomplete knowledge of it. Using probability distributions and how these probability distributions transform in time or in order to be able to predict the behaviour of a system does not entail that there is an inherent chance in the heart of the system, rather it implies that our knowledge of the system is incomplete and limited\footnote{ The probability function should follow Kolmogorov axiomatization Hájek, A. (Winter 2012 Edition). Interpretations of Probability. The\ Stanford Encyclopedia of Philosophy  E. N. Zalta. }. \par

\vspace{\baselineskip}
\textit{Predictability}: Predictability is an epistemological category concerning our ability to predict the state of a system at all points once we know its state at a given point. Suppose that we have an observer and a system that is isolated from its environment, predictability is the epistemological category concerning the ability of the observer to predict the state of the system at all given point once the observer knows all its initial conditions at the given point.\par

\vspace{\baselineskip}
\textit{Chance:} We define chance as some Epicurean ontological chance (the way Leibniz understands Epicurean chance in his correspondence with Clarke), i.e. the system is governed by genuinely stochastic and indeterministic laws\footnote{ Formally, it is defined by Lewis’ Principle Eagle, A. (Spring 2019 Edition). Chance versus Randomness. The Stanford Encyclopedia of Philosophy. E. N. Zalta.. }. \par

\vspace{\baselineskip}
\textit{Randomness:} We define randomness as unpredictability. Formally\footnote{ The mathematical definition of randomness was first formalized by Von Mises in 1919 and then Mises–Wald–Church. } randomness is defined with respect to a given class of sets. ‘Randomness is a relative notion, not an absolute one, in that it depends on the choice of the set of selection rules’ (Eagle Spring 2019 Edition). The mathematical definition of Randomness relates to computation theory which proves that an absolute random number is not computable and can never be generated\footnote{ This results from Gödel’s famous incompleteness Eagle, A. (Spring 2019 Edition). Chance versus Randomness. The Stanford Encyclopedia of Philosophy. E. N. Zalta. }.\par

\vspace{\baselineskip}
\textit{Causality}: We take Bell’s Principle of Local Causality which emphasizes that no information can travel faster than light. Bell formulated this principle as ‘the direct causes (and effects) of events are nearby, and even the indirect causes (and effects) are no further away than permitted by the velocity of light.’(Myrvold Spring 2019 Edition)\par

\vspace{\baselineskip}
Now that we have settled on some key definition, we will reflect on Chaos theory and employ it to better understand Leibniz’s philosophy. We will only give a very brief reflection on the interpretation of Quantum Mechanics in the epilogue, but further discussion of that will be beyond scope of this article, and we limit this article to the discussion around Chaos Theory. A connection between Leibniz’s philosophy and Chaos theory has been already suggested by Bouquiaux (Bouquiaux 1993)\  who found these two ‘startlingly similar’. In this article, we suggest that Chaos theory in the way that it is understood by physicists like Jean Bricmont (Bricmont 1995) can be enlightening for this goal. Jean Bricmont understands Chaos theory through his reading and interpretation of Laplace’s philosophical essay on probabilities (Marqui 2013). In this article, we will attempt to establish a more rigorous connection between Chaos Theory and Leibniz’s theory which preceded Laplace. We especially focus on Leibniz’s correspondence with Clarke (Leibniz, Clarke et al. 2000). These correspondences are an epitome for utility and the necessity of avoiding a dichotomy between science and philosophy.\par

\vspace{\baselineskip}
It is our position that philosophy should again be revised without the premise of the above dichotomy, as many scholars have sought to bridge this gap. This article argues that aside from misleading and sometimes inaccurate scientific metaphors of some French philosophers which provoked the Science war, we need to bring the scientific faculty back to philosophy, and devise a comprehensive system of thinking like inspired by the philosophy of Leibniz. We believe that today both philosophy and social science should be developed and expressed more closely with science. Hence, a closer collaboration between scientists and philosophers/social scientists is advocated here. Philosophical concepts of Chaos theory will be reframed to shed new light on concepts of determinism. \par

\vspace{\baselineskip}
\textbf{\textit{II. Determinism and the rationalistic tradition of Leibniz}}\par

\vspace{\baselineskip}The dualism of Descartes has a serious problem, or rather two problems that are two sides of the same coin: the problem of cognitive grounding addressed by Fichte and Schelling and the problem of the body and mind relation addressed by Pierre Gassendi who was contemporaneous with Descartes in theology (Duncan 2016). The big question is how in a linear causally closed physical world of appearances determined by the laws of physics, the cognizant subject can have free will? If objects in the outside world of the cognizant subject are only a collection of ideas, and ideas are the only true eternal immortal being, how then does the cognizant subject assure oneself at all that there are \textit{thing-in-themselves }that provide for the ground for his representation of the existing world? Humans as a logo-centric cognizant subject question their relation to the world outside of oneself. This questioning gives rise to a contrast in which one differentiates oneself from the outside world so that one sees one's very being as the faculty of reason that gives one an identity; a self, an I, which is separate from the world. By this approach, nature is reduced to the status of being a mere object of thought, and by doing so he himself becomes the subject of his own thoughts. Hence, by coming to deny the outside world that lies before him, turning the locus within and equating being with thinking, he loses any grounding for existence. In consequence, ‘I think’ does not strongly lead logically to ‘I am’.\par

\vspace{\baselineskip}The discussion on the relation of nature and free-will in entangled with the question of necessities and contingencies. The special distinctive aspect of this contention was that in light of the new physics, theological discussions on this matter were transformed into scientific discussion. With Leibniz on one side and Clarke and Newton on the other side, this brought the theological argument among Franciscans and Dominicans into a whole new domain of physics (Redding Fall 2011). One important incident that especially gave us insight into the double sidedness of this dispute is the correspondence between Leibniz and Clarke during the last months of Leibniz’s life (Leibniz died before being able to answer Clarke fifth letter addressed to him). In these correspondences (Leibniz, Clarke et al. 2000), Clarke questions Leibniz’s idea of pre-established harmony and objects that not only does Leibniz’s determinism leave no room for human free will, but also no room for the free will of God to exercise His power and to perform miracles. Clarke believes that miracles need no cause other than the ‘mere will’ of God. Leibniz answers that ‘I maintain that God has the power of choosing since I ground that power on the reason of a choice agreeable to his wisdom’(Leibniz, 2000, 3:8, 16). And he proceeds in his fifth letter: ‘God’s perfection requires that all his actions should be agreeable to his wisdom and that it may not be said of him that he has acted without reason.’ (Ibid, 5:19, 39). For Leibniz, God’s wisdom has primacy over His power, and God’s will follows only His reason, i.e. God’s wisdom is the same as His will.\par

\vspace{\baselineskip}Miracles have been contentious in all theological debates; however during the 17\textsuperscript{th} and 18\textsuperscript{th} centuries with the perspective of new scientific determinism miracles became more problematic for theologians. A coherent, rationalistic and deterministic picture of the world as a whole was formulated by physics, and for the first time, the whole universe seemed explicable. With this view of the world, the question of what a miracle is and how it is performed in nature was raised. Do miracles that happen in the physical and material world defy physical laws? Leibniz writes ‘If God has to mend the course of nature from time to time; he must do it either supernaturally or naturally. If supernaturally, then this is appealing to miracles in order to explain natural things yet that amounts to a \textit{reduction ad absurdum} of this hypothesis, for once you let in a miracle anything can be explained with no trouble at all.’ (Ibid, 2:12, 11) Questioning miracles led to a far more important fundamental question on the causality of the physical world, and the decisiveness and determinism of physical laws. In their argument Leibniz and Clarke both show inflexibility and a degree of prejudice toward their theological faith as well as their understanding of physics, and in their heated debate, they both end up somehow giving a definition to a miracle that is both fluid and subjective. Clarke ends up defining miracles as things that are ‘unusual’ to us, whereas Leibniz objects that ‘the monsters [human or other animals that have from birth physical features making it strikingly unlike most members of its species] would be miracles;’ (Ibid, 4:43, 27) and Leibniz thinks of miracles as things that cannot be explained by the laws of physics. This resulting fluidity of the conception of the miracle is raised due to the unanticipated overlapping of theology with natural science, and as a further result of this fluidity, a final and clearer explanation of the miracle becomes impossible. Leibniz starts his argument under the assumption of perfection of the natural world, and writes in his first letter: ‘I hold that when God works miracles, he does not do it in order to supply the needs of nature, but those of grace.’ (Ibid, 1:4, 4). By this, Leibniz insists that miracles should not be explained supernaturally and speculates that miracles are phenomena that cannot be explained by natural laws of his time, but they also should follow some physical laws unknown to them. In his third letter, he writes, ‘The supernatural exceeds all the power of created things.’ (Ibid, 3:17, 17). And goes on to say that a miracle is something that ‘cannot be explained by the nature of bodies [i.e. natural laws of physics].’ Is he contradicting himself? If a miracle follows deterministic laws of nature, it would then not be a miracle, but a natural thing, a part of nature. And if a miracle is a result of the grace of God, and not a product of nature coming by way of the supernatural, then this contradicts the assumption of the perfection of the world, i.e. ‘God can only choose what is best $ \ldots $  For what is necessary is so by its essence, since the opposite implies contradiction; but a contingency that exists owes its existence to the principle of what is best, which is sufficient reason for the existence of things.’ (Ibid, 5:9, 37).  In common understandings of miracles which Clarke put forward, miracles are intervention and an amendment by God in the natural course of events. For Newton, nature is a watch that needs God’s amendment from time to time, nature is constantly losing force, i.e. energy, and God is the source of this energy; so God’s amendment means God is acting as the source of energy. Leibniz with his idea of a perfect world and pre-established harmony opposes this, and instead puts forward for the first time in the history of physics, the Law of Conservation of Energy. Leibniz thought of the natural world to be so perfect that any essential disorder or chance was not allowed. So for Leibniz, a miracle is just a human concept for things that he cannot explain by his limited knowledge of the natural world. What humans call a miracle is not an intervention or amendment of God in the natural course of events but it is part of nature that is unknown to us. In this correspondence, one main problem discussed is Newton’s law of gravitation, i.e. how the earth is orbiting around the sun. At that time physics only meant Impact Mechanics, which is the physics of colliding bodies, so Leibniz attacks Clarke’s criticism of Newton that he only provides a formula for gravitation and yet has not explained the mechanism of gravitation. Leibniz writes:\par

\vspace{\baselineskip}\begin{adjustwidth}{0.49in}{0.0in}
I objected that an attraction [of celestial bodies] properly so-called would be an operation at a distance without any means intervening $ \ldots $  What does Newton mean when he will have the sun attract the globe of the earth through empty space? Is it God himself that performs it? But this would be a miracle if ever there was any. Or are perhaps some immaterial substances of some spiritual rays, or some accident without a substance, or some kind of intentional species, or some other I-know-not-what, the means by which this is claimed to be performed? Of these sorts of things, Clarke seems to have still a good stock in his head, without explaining himself sufficiently. The means of communication, he says, is invisible, intangible, and not mechanical. He might as well have added inexplicable, unintelligible, precarious, groundless, and unprecedented. But it is regular, Clarke says, it is constant, and consequently natural. I answer that it cannot be regular without being reasonable, nor natural unless it can be explained by the natures of creatures [laws of physics]. If the means which causes an attraction properly so-called are constant and at the same time inexplicable by the powers of creatures [laws of physics] and yet are true, it must be a perpetual miracle, and if it is not miraculous, it is false. (Ibid, 5:118, 64)\par

\end{adjustwidth}

\vspace{\baselineskip}In his fifth letter which Leibniz was ‘prevented by death from returning any answer’, Clarke responds to this:\par

\vspace{\baselineskip}\begin{adjustwidth}{0.49in}{0.0in}
The phenomenon itself, the attraction, gravitation, or tendency of bodies toward each other, and the laws or proportions of that tendency are now sufficiently known by observations and experiments. If Leibniz can explain these phenomena by laws of mechanism, he will not only not to be contradicted, but will moreover have the abundant thanks of the learned world. (Clarke, 2000, 5:124, 86)\par

\end{adjustwidth}

\vspace{\baselineskip}It is only today that physicists can explain this mechanism after the theory of General Relativity. Physics of Leibniz’s era could not explain this phenomenon; this inability of their impact mechanics to explain the cause of gravitation gave room for the use of the concept of the miracle. This conception of miracle provided for a strong decisive determinism in nature which through its incompatibility with human knowledge opens a door for avoiding this very determinism in the human sphere altogether.\par

\vspace{\baselineskip}\begin{justify}
From certain writings of Leibniz, it might seem that he believed in two separate domains, that of nature and of ethics, the domain of necessity and the domain of freedom, Leibniz uses two different words, \textit{nature} and \textit{grace}. In his first letter to Clarke he writes ‘when God works miracles, He does it not to meet the needs of nature but to meet the needs of grace’, in his fifth and last letter to Clarke, he writes ‘all the natural forces of bodies are subject to mechanical laws, and all the natural powers of spirits are subject to moral laws. The former follows the order of efficient causes, and the latter follows the order of final causes.’  It is as if for Leibniz there are two worlds each with its own laws and that Leibniz before Kant believes that humans live in two incompatible worlds. In an important treatise Leibniz wrote in 1714, \textit{principles of nature} and \textit{grace based on reason}, the same year he wrote \textit{Monadology}, he speaks as if nature and grace are separated, he speaks of \textit{kingdom of nature }and \textit{kingdom of grace} which can imply two separate domains, Kant employs this very distinction in his \textit{Critique of Pure Reason (Immanuel Kant and Paul Guyer (Editor 1999)}. However, we think that Leibniz did not believe in two separate worlds with separate and incompatible laws. A more careful study of this text does not support this interpretation. Leibniz emphasizes that ‘supreme wisdom and goodness can only act with perfect harmony’, he invites us to listen to diverse voices in the world singing in harmony. The separation between the domain of \textit{nature} and \textit{grace} is devised within the context of his theory of pre-established harmony which implies there is no separation at all. He writes ‘nature itself leads to grace, and grace perfects nature by making use of it’. His principle of pre-established harmony means that the actual world is the best possible world in its design which although it is maximum diversity and complexity, it is at harmony. ‘God chooses the best possible plan in producing the universe, a plan in which there is the greatest variety with the greatest order.’ (Ariew and Garber 1989) In his correspondence with Huygens about the same as his correspondence with Clarke, he writes ‘everything happens mechanically in nature, but that the principles of mechanism are metaphysical, and that the laws of motion and nature have been established, not with absolute necessity, but from the will of a wise cause, not from a pure exercise of will, but from the fitness of things.’ 
\end{justify}\par

\vspace{\baselineskip}
\begin{justify}
Our interpretation is that Leibniz extends laws of nature in a complex network of free agents in such a way that he aims at giving a coherent picture of nature that is complex. The simplistic and reductionist interpretation of this would be to say that there is no free will. However, for Leibniz, the universe has a sufficient reason, i.e. \textit{causa finale}. For Leibniz, freedom means acting upon \textit{causa finale} of the universe. In his fifth letter he writes ‘It is true that according to me the soul does not disturb the laws of the body, nor the body those of the soul, and that the soul and body do not agree together, the one acting freely according to the rules of final causes and the other acting mechanically according to laws of efficient causes. But this does not derogate from the liberty of our souls, as the author here [Clarke] will have it. For every agent who acts according to final causes is free, although it happens to agree with an agent acting only by efficient causes without knowledge, or mechanically, because God, foreseeing what the free cause would do, did from the beginning regulate the machine in such manner that it cannot fail to agree with that free cause.’ Therefore freedom means getting more knowledge by way of getting closer to the mathematics of the world. This means a rationalistic view of the world in which as Leibniz writes in \textit{principles of nature} and \textit{grace based on reason}, ‘plan in which there is the greatest variety together with greatest order$"$  (Leibniz, sec: 10 Prin, 210)
\end{justify}\par

\vspace{\baselineskip}\begin{justify}
Leibniz's attempt to dissolve Cartesian dualism came to naught with Kant bringing back this dualism with concomitant subjectivism, i.e. an irreconcilable gap between subject and object, between free cognizant subject and the world of phenomena. In a dualism of nature \textit{in itself} and nature as an \textit{appearance}, there is a risk of knowledge becoming subjective. He puts the mechanism of the natural world against the free human world. Kant leaves human judgments, such as aesthetic or ethical judgments, to a non-physical realm. The irreconcilable rationalistic dichotomy between appearing nature and nature \textit{in itself} has its roots in the simple deterministic view of nature akin to Newtonian physics. There are two interchangeable ways to resolve this dualism. One is to abandon the idea of \textit{things-in-themselves} altogether, saying that as \textit{things-in-themselves} cannot be known and have no interaction with each other, so they should necessarily be \textit{indeterminate}. Therefore, they cannot get any \textit{determinate} form or identity in themselves, and subsequently, any attempt to identify any transcendent \textit{thing-in-themselves }is in vain. The second way is to abandon the inherited Cartesian dualism of mind/matter in Kant’s philosophy. The dualism of Descartes and Kant jeopardize the entirety of scientific faculty and a return to Leibniz's philosophy seems imperative to us.
\end{justify}\par

\vspace{\baselineskip}A view of total subjectivity promoted by some phenomenologists can never be scientific nor achieved. Science always has an objective claim on knowledge. Scientific claims can be refuted, but their capability to be falsified does not invalidate their claim to objectivity. Consider two scenarios; first, our best claim on any objective truth is at the level of a warranted belief, in which case scientific claims have the highest level of objectivity since they are evidential. Second, there exists some universal truth beyond our knowledge of sufficient evidence, in which case the best image one can have of that truth is the evidential view of science. In either case, it should be possible to move beyond a self-evident belief or opinion. Although knowledge, as discussed by some, is self-referential, it is not self-evidential. The work of scientific knowledge acquisition is not to construct evidence for a theory, but rather a verification of a theory with the help of empirical evidence. It is too naïve to view science in light of the first position. And in this light, it is nonsensical to abandon the objective approach of science altogether and go for a totally subjective view.\par

\vspace{\baselineskip}\textbf{\textit{III. Determinism in Chaos theory}}\par

\vspace{\baselineskip}Considering that our problem in this article is the relation between man and nature, we recognize two levels of this relation; on a pragmatic level, within the scale of human experience, that is larger than a molecule size and smaller than cosmological scale Laws of Classical Mechanics hold true. In this scale, for dealing with everyday needs of humans, Chaos Theory and Statistical Mechanics which are classical theories suffice for explaining the relation between man and nature. One can argue that today our daily experience is radically changed by technologies that are governed by laws of Quantum Mechanics, and furthermore on a more fundamental philosophical level, where ontological and epistemological questions concerning the relation of man and nature are asked, fundamental theories like Quantum Mechanics pose a significant problem in our understanding of the world. The problems of determinism rose in the twentieth century modern physics in a whole new fashion with a rupture in the physicists’ coherent picture of the world, with two grand theories, General Relativity and Quantum Mechanics, first appeared to be incompatible exactly because of the seemingly inherent randomness of Quantum Mechanics. One can argue that Chaos theory is a classical theory that deals mathematically with a limited set of physical phenomena, and it is not a fundamental theory like Quantum Mechanics and General Relativity. Yet, Chaos theory works within a framework of classical physics and is at our focus in this article. Discussion over fundamental theories like Quantum Mechanics and General Relativity is beyond the scope of this article. However, at the end of this article in an epilogue, we briefly give our opinion on the fundamental theories, especially rejecting pragmatic Copenhagen interpretation of Quantum Mechanics. Let us turn our focus on Chaos Theory now.\par

\vspace{\baselineskip}Chaos, i.e. Nonlinear Dynamics, however, does not entail an ontological chance, Chaos is an \textit{orderly disorder.}\  In view of Chaos Theory, the reality exists independent of\ the observer.  The world is not mere mental entities, but unlike the viewpoint of nineteenth-century science, humans cannot grasp it wholly and completely. In Chaos, causes are always over-flowing, tipping from one domain to another domain. In a complex system, the emergent properties are interdependent and irreducible to a linear chain of cause. In Chaos Theory, it is famously said that flaps of a butterfly in Argentina may cause a tornado in Texas three weeks later. A small change in a system could lead to a large change in system behaviour which is known as \textit{sensitive dependence }(Anishchenko 2007). There is a \textit{path dependency} in which causation flows from contingent events to general processes, from small causes to large system effects, and from historically or geographically remote locations to the universals. \textit{Path dependency} shows that a small fluctuation in the processes through time significantly influences the effects that might turn out a long time ahead of the process. It should be noted that Henry Poincare and Robert Battermam emphasized\  the problematic definitions of Chaos stating that exponential instability, i.e. the exponential divergence of two trajectories issuing forth from neighbouring initial conditions, also known as \textit{sensitive dependence }on initial condition, is a necessary condition, but it is not a sufficient condition for Chaos to happen (Bishop 2017). Determinism in Chaos Theory means that there exists a real world out there independent of the observer with the deterministic Laws of Physics governing it, but it does not entail that the outcome of a physical system can be fully and completely determined. The world is causal, but its causality is not linear; i.e. A simply resulting into B. \par

\vspace{\baselineskip}The lesson to learn here is to look at nature firstly not as a linearly closed system of causation; causes A simply effecting B. And secondly, not to think of natural systems run by chance in which anything can happen. Chaos is different from blind chance. Chaos is not linearly causal, but there is a certain amount of coherency between different incidents i.e. events are \textit{correlated} in a complex chain of causes and are not merely random. The problem for many philosophers in understanding Chaos theory, as rightly exposed by Jean Bricmont, is that they reduce Chaos to the \textit{Butterfly effect, }i.e. \textit{sensitivity dependence}, while this is only a necessary condition for Chaos and not a sufficient one. The output of Chaos is not totally random and shows coherency. Chaotic phenomena are ‘governed by deterministic laws, and therefore predictable in principle, which is nevertheless unpredictable in practice because of their \textit{sensitivity to initial conditions}’ (Sokal and Bricmont 1998). Although \textit{sensitive dependence} on initial conditions is a necessary aspect for defining chaos, for an understanding of whether a system has just arbitrary behaviour, or indicative of deterministic chaos, other criteria, including certain \textit{topological characteristics},\textit{ }should also be met. \par

\vspace{\baselineskip}
In the heated controversy of the Science War a close collaboration between physicist Ilya Prigogine and philosopher Isabelle Stengers took place (Prigogine, Stengers et al. 2018), they tried to incorporate Chaos Theory into their philosophy. However, Jean Bricmont (Bricmont 1995) rightly denounced their misunderstanding of Chaos Theory. Bricmont (Bricmont 1995) criticizes Prigogine and Stanger for thinking that ‘The existence of chaotic dynamic systems supposedly marks a radical departure from the fundamentally deterministic worldview, [and] makes the notion of trajectory obsolete.’ He writes: \par

\vspace{\baselineskip}
\begin{adjustwidth}{0.5in}{0.0in}
\begin{justify}
Chaos does not invalidate in the least the classical deterministic worldview, the existence of chaotic dynamical systems actually strengthens that view$ \ldots $  I think this is based on a serious confusion between determinism and predictability. In a nutshell, determinism has to do with how Nature behaves, and predictability is related to what we human beings are able to observe, analyze, and compute$ \ldots $  Anybody who admits that some physical phenomena obey deterministic laws must also admit that some physical phenomena, although deterministic, are not predictable, possibly for ‘accidental’ reasons. But once this is admitted, how does one show that any unpredictable system is truly nondeterministic, and that the lack of predictability is not merely due to some limitation of our abilities?
\end{justify}\par

\end{adjustwidth}

\vspace{\baselineskip}\begin{justify}
We can trace back this incompleteness of our knowledge to the beginning of the Universe. In Statistical Mechanics, it is contentious whether it is possible to deduce irreversible macroscopic processes through statistical calculations solving microscopic dynamical laws of physics for large numbers of particles, which are time-symmetric and reversible. Imagine an insulated container of gas separated into two chambers with a moveable heat-conducting membrane in between so that two chambers are at the same temperature, one side contains air and the other side a different gas, say argon. An experimenter takes away the membrane so that the gases will diffuse throughout the volume; you intuitively expect that the gases mix, that the argon in one-half diffuse to the entire volume, the Second Law of Thermodynamics also dictates this. And it is this very irreversibility that gives a sense to the Arrow of Time. At the end of the 19th century, L. Boltzmann formulated a derivation of this process, hence the Second Law of Thermodynamics, from statistical derivation solving for Mechanical laws governing a large number of particles of the gas. The law of Impact Mechanic for these particles is symmetric, meaning that if we reverse the direction of velocities, i.e. solving for vectors of velocities in collisions of these particles, the law of Impact Mechanic still holds true. The objection was raised by J. Loschmidt (Callender Winter 2016 Edition) that if reversing of all velocities of all particles is allowed, which law of physics stops us from thinking a scenario where all particles of the argon gas will proceed to gather in one half of the container\footnote{ Loschmidt showed that Boltzmann's assumption that velocities of all these particles are uncorrelated cannot be true. }, which laws of physics can determine the direction of time? Jaynes (Jaynes 1965) famously challenged the formulation of Boltzmann and argued that Entropy, that is an increase of disorder in the system to which we assign the direction of time, is ‘anthropomorphic’. Bricmont (Bricmont 1995) objects and respond ‘A better word might be ‘contextual’ i.e. they depend on the physical situation and on its level of description.’ He defends Boltzmann’s formulation and answers to these objections. He writes:
\end{justify}\par

\vspace{\baselineskip}
\begin{adjustwidth}{0.5in}{0.0in}
\begin{justify}
The only real problem with irreversibility [as stated in the second law of thermodynamics] is not to explain irreversible behavior in the future, but to account for the exceptional initial conditions of the universe$ \ldots $  If the laws [of physics] are deterministic, assumptions on initial conditions are ultimately assumptions on the initial state of the Universe. Once one has remarked that there is no contradiction between irreversibility [of macroscopic physical phenomena] and the fundamental laws [of physics], one could stop the discussion. It all depends on the initial conditions. 
\end{justify}\par

\end{adjustwidth}

\vspace{\baselineskip}
\begin{justify}
While classical statistical mechanics does not represent the overall behaviour of the systems considered because their great mechanical complexity prevents such a representation, it assumes that the individual constituents of these systems are represented by classical mechanics. Indeed, one could experimentally isolate, say, a molecule of gas and make deterministic predictions concerning its behaviour. In many cases these causal processes, even in dealing with more complex systems, can still be treated by linear equations, \ but in other cases  with more complex causal chain require nonlinear equations. A complex system of overflowing linear causes can be analysed as a nonlinear system. In Chaos Theory, \textit{Irreversible path dependence} might occur over time when contingent events set into motion patterns which themselves have deterministic properties (Urry 2005), and does not entail the system is inherently run by chance. For Bricmont, ‘This just reflects the fact that different points in the support of the initial distribution, even if they are close to each other initially, will be separated by the chaotic dynamics’ (Bricmont 1995). 
\end{justify}\par

\vspace{\baselineskip}
To be able to explain \textit{Irreversible path dependence} and distinguish it from random occurrences, we need to explain Chaos theory in more technical terms. In Nonlinear Dynamics, the terms \textit{Phase space}, abstract mathematical space where each point represents each state of the dynamical system, gives the unfolding of a chaotic system at a given point in time.  Mathematically, the state of the system having multiple degrees of freedom is represented by a point in its \textit{phase space}. Each point in that space represents the positions and the velocities of all the particles of the system under consideration. \textit{Phase trajectories }are the trajectory of the system from one state to another. \textit{Phase space }is space of all possible \textit{Phase trajectories} and encompasses different features, namely \textit{limit cycles}, \textit{attractors},\textit{ repellers} and \textit{saddle points}. \textit{Phase trajectories }tend towards \textit{attractors}. A \textit{saddle} point (or surface) occurs when some of the phase trajectories enter and others leave the \textit{saddle point} (some are pointed inward and others outward), in which the system experiences an extreme transformation, usually associated with unstable potential functions\textit{. }Some \textit{attractors}, namely \textit{strange attractors,} can have an intricate geometrical structure.\footnote{ Strange attractors can be illustrated in a Horse Shoe map. A required condition for chaos to occur, according to Batterman, is the existence of a stretching and folding mechanism which makes some trajectories converge and some diverge rapidly. Along with such conditions to exist for a chaotic system, an aperiodic situation is produced, which means they present no repetition of themselves on any time scales. In Smale’s map, when the procedure of stretching and folding is repeated many times, in the limit, the original square gets mapped into a series of strips which have Cantor structure in its cross-section. Anishchenko, V. S. (2007). Nonlinear dynamics of chaotic and stochastic systems. Berlin ; New York, Springer: xvi, 446 p. } One of the significant features of \textit{strange attractors} is having a \textit{self-similar} structure which is repeated on arbitrarily small scales. This kind of \textit{attractor} enjoys an infinite number of layers of a repetitive structure. This quality makes trajectories linger within a bounded region of state space by folding and intertwining with one another without ever intersecting or repeating themselves. A closed trajectory in phase space on a manifold is called a \textit{limit cycle} (\textit{Van Der Pol oscillator} for example). \textit{Bifurcation} caused by the variation of system parameters bringing structural rebuilding of \textit{Phase portrait} and breaking the earlier topological equivalence. \par

\vspace{\baselineskip}\textbf{\textit{IV. Refuting dualism of natural and human laws}}\par

\vspace{\baselineskip}Any discussion on the relation between natural science and human sciences will come back to an understanding of the relation between man and nature. The proposed concepts for the relation between these two so far include harmony, correspondence, separation, disjunction, opposition, adversary, dependence, functionality; all based on a dualistic approach, be it absolute or relative, or on a reductionist approach. Here, for overcoming this dualism we propose the concept of complexity, not in its general meaning, but rather in a very specific meaning formulated by Chaos Theory. We suggest here that even though it is not possible to completely avoid this conundrum in day-to-day language in a satisfactory way, as natural language and mentality are inclined towards this dualism, toward making the ‘I’ as the cognizant subject and the world as the object outside of this very cognizant subject, as the dualism of man and nature is at its core our most innate experience of the world, the mathematical model of complexity, however, formulated in Nonlinear Dynamics is capable of transcending this dualistic approach. In our rendition, all that there is in the world is complex, and complexity in a seemingly contradictory way makes up a whole that we call the world. We here think of the world as an \textit{a priori }closed whole, but as an \textit{a posteriori }open world, which complexity is at play in a very large complex network. This complexity is such that a lot of things that look like an impasse to us, and come as miracles, are resolved not through a transcendent intervention, but rather are possible through immanent natural contingencies of the complex system itself. The domain of possibility in a complex system is far broader than our historical measurements, estimations, and predictions, but it is not hap hazardous. The very opposing forces of the system itself can give room to multiple possibilities. The more information we have about the system, the clearer a picture we have of the system and the better we can predict the future of the system, but we can never fully grasp it and predict it in its full certainty. The structure of human understanding is such that it tends toward making things simple and gives simple explanations based on what one knows so that one assures oneself that he has come to know the thing fully and completely. In our rendition, we recognize the principle of discernibility, that is a man can discern and acquire knowledge of things, but we suspend the notion of the complete realization of such knowledge. We are sceptic in so far as the totality of human knowledge of the system is in question, but we alleviate the system from our skepticism and we grant an ontological wholeness and an ontological determinism for the system itself. It is exactly because of our incomplete knowledge of the system that we grant at the same time open possibilities in the system. We recognize causality in the world and as a result, the possibility of exact science, but we do not render causality to be linear, thus no science can have a complete final picture of the world because of the limited span of history and the limited knowledge of its initial conditions.\par

\vspace{\baselineskip}As far as the relation of this duality with Leibniz‘s philosophy and its relation to Chaos Theory, Bouquiaux believes the spirit of Leibniz’s philosophy of nature ‘startlingly similar’ to today’s Chaos theory (Bouquiaux 1993). We claim that not only is Leibniz’s philosophy ‘startlingly similar’ to Chaos Theory, but Leibniz’s ontological system is in line with determinism in Chaos Theory.  In \textit{Monadology}, Leibniz made a major contribution in suggesting interminable finer graining processes which are akin to processes in Chaos Theory. Leibniz reflects on the relationship between finite and infinite and how finite covers infinite. Any particle in his world, although small, includes and expresses the infinite \textit{in itself}. In \textit{Monadology, }Leibniz writes: ‘Every portion of matter is not only divisible to infinity but is actually sub-divided without end... Without this, it would be impossible for each portion of matter to express the whole universe’ (von Leibniz and Latta 1898). Inevitably, this led him to the question of necessity and \textit{contingency}. Leibniz believes in two modes of truth; the truth by \textit{necessity} and the truth by \textit{contingency}. For him, a true proposition is the one whose predicate is included in the subject itself. A necessary proposition can be simplified to identity i.e. its contrary leads to a contradiction. But when proving a proposition continues to the infinite (\textit{resolutio procedit in infinitum}), this proposition is called \textit{contingent}. Necessity and \textit{contingency} are like real and irrational numbers. A proposition that is true by necessity turns into identity, but contingent truth, like irrational numbers, is infinitely divisible. In any true proposition, the predicate is included in the subject, but in the contingent truth, it is \textit{virtual}, only known as \textit{a posteriori} i.e. it is given in an implicit way like in the number $ \pi $ . Leibniz sees the structure of nature in this way, each part includes the whole. A \textit{representation} is not something separate from what is represented. Representations are genuinely internal, not a window from out to in, or vice versa. Like for Leibniz, the predictability of contingent truth is only an ontological \textit{a priori }while in practice it is \textit{a posteriori }and any truth can be unpredictable due to our lack of knowledge about initial conditions. Determinism depends on what Nature does independently of us, while predictability depends in part on Nature and in part on us. Complex system theories do not follow a simple causally closed set of rules, but the output is not arbitrary and run by chance, i.e. one should not give in to a total subjective relativistic view.\par

\vspace{\baselineskip}
\textbf{\textit{V. Conclusion}}\par

\vspace{\baselineskip}This article is an attempt to introduce a relation between science and philosophy. Our final goal will include a contribution to art theory and its disposition in society. Social incidents, like in Chaos, are not simply linearly and causally related, in a sense of a linear continuity of history, but historical coherency can be witnessed in them. Let us go back to what seems nature’s erratic irregular behaviour, say, how an earthquake can be explained? Someone with a mythological mentality would respond that it is the work of God, while a scientific mind will attempt to find a rational explanation for it. Here lies the clue for a better understanding of nature and for avoiding the dualism. A modern scientist would not say it is the work of God nor run by chance, a mystery, but instead would say it is the outcome of a very complex system governed by a network of deterministic laws. The outcome of the system is not strictly determined because we cannot have enough information about the conditions of a system. A scientist is never afraid of updating his understanding of a Causal chain. A scientific mind does not take the world as an accumulation of random events, nor does he believe in the work of mysterious agents. A scientist seeks a rational cause and effect pattern in every phenomenon, but he is not afraid to revise his most basic assumptions and most basic principles about the world. And for the twentieth-century scientific mind cause and effect is not a simple linear A to B, but rather overlapping chains of causes within a complex system. Social incidents, like in Chaos, are not simply linearly causally related. \par

\vspace{\baselineskip}In the pre-modern era, the relation between science and philosophy was such that science conformed to philosophy. Since 17th-century various wars between science and philosophy changed this relation in different ways: a complete separation of the two by means of denouncing one and affirming the other; an absolute separation of two by accounting for two mutually exclusive domains and affirming each in these separate domains; one following the other. The last but not  least, a fundamental correspondence between the two in principle and ends. In this article, we adhere to this last alternative, for which we trace traits of this correspondence in physics and philosophy of Leibniz. In Leibniz’s philosophy correspondence between philosophy and science, and thus a correspondence between natural science and humanities, is established by demonstrating inevitable certainty and determinism of fundamental laws that operates in the natural realm as well as in the human realm. In this article, we employed both literatures of philosophy as well as physics to defend this ontological determinism of fundamental laws. The natural consequence will be that the world is a realm of necessity and determined universality which leaves no room for randomness. But then how would we be able to think of free will which is the basis of assigning value to human actions? The dominant philosophical tradition since Parmenides is that, by adhering to determinism, movement, and multiplicity, and as a result, free will is deemed to be mere illusion. Kantian tradition separates the human realm from the natural realm and renders any relation and correspondence between philosophy and science impossible, we avoid this route. To our belief, the statement ‘the world is necessary’ is true from an ontological perspective and the statement ‘in the world free agents exist’ is true from an epistemological perspective. We have to, however, be attentive not to issue yet another dualism, and that is why we use the concept of ‘lack of knowledge’. The problem is not to say because human knowledge works in another realm so we can claim free will for certain beings, but the problem is that because the cognizant subject is faced with a relative lack of knowledge in the necessary world, he acts only relying on available data. His actions, as they are not based on knowledge of all needed data, are called actions of free will. In this way, free will is an epistemological translation of this ontological statement that a ‘certain being, i.e. cognizant subject, has never access to all information, and can never fully and completely grasp what is there’. In modern physics, Chaos Theory demonstrates well this very statement, it states that the world is deterministic but the cognizant subject can never fully and completely grasp the whole. Should we give in to a fatalistic view of the world saying that free will is a mere illusion? The answer is no. The answer to this question is however super complex and requires infinite endless elaboration.\par

\vspace{\baselineskip}\textbf{\textit{VI. Epilogue }}\par

\vspace{\baselineskip}The reader might contend that Chaos Theory is Classical theory and it is not a fundamental theory like Quantum Mechanics and General Relativity, which deal with the ultimate constitution of nature and fundamental forces of nature. Carlo Rovelli (Rovelli 2007) argues against the anti-philosophical ideology of, especially prestigious physics Steven Weinberg and Stephan Hawking. Weinberg wrote a chapter titled ‘Against Philosophy’ in which he argues ‘philosophy is more damaging than helpful for physics’, and Hawking famously wrote that ‘philosophy is dead’ as he believed today ontological questions are dealt with by physicists rather than philosophers. Rovelli believes this attitude has ‘damaging effects on the fertility of science’, he believes ‘philosophy has always played an essential role in the development of physics’. He writes: \par

\vspace{\baselineskip}\begin{adjustwidth}{0.43in}{0.0in}
‘Weinberg, Hawking, and other anti-philosophical scientists are in fact paying homage to the philosophers of science they have read, or whose ideas they have absorbed from their environment. Thus, when Weinberg and Hawking state that philosophy is useless, they are actually stating their adhesion to a particular philosophy of science. In principle, there’s nothing wrong with that; but the problem is that it is not a very good philosophy of science, Popper, and Kuhn, so popular among theoretical physicists, have shed light on important aspects of the way good science works, but their picture of science is incomplete and I suspect that taken prescriptively and uncritically, their insights have ended up misleading research’ (Rovelli 2018)\par

\end{adjustwidth}

\vspace{\baselineskip}In this article, Leibniz’s philosophy served as a link between science and philosophy. Furthermore, we believe his philosophy can help to relieve some confusion in both domains. The coherence of Leibniz’s philosophy with determinism in Chaos theory was shown. But, what Chaos theory has to do with our new understanding of the world? On one hand, on a technical level, Nonlinear Dynamics is at work both at the advent of the universe is rooted in an infinitely small difference that has been extended to complicated differential fields as well as near Black Holes. General Relativity is by nature nonlinear, as explored by Wheeler (A.Wheeler 1957). Quantum Mechanics, although first modelled in a linear model, has matured into the nonlinear Schrodinger equation since the 1950s (Ablowitz 2008). On the other hand, on a more conceptual and fundamental level, which is important to us here, Chaos Theory is fundamentally deterministic. The reason that we cannot strictly determine the outcome of a physical system is our lack of knowledge and not because the system is fundamentally governed by blind chance. And Bohmian mechanics is also ontologically deterministic. \par

\vspace{\baselineskip}
One might argue that Quantum physics does not help us with the problem of free will as it deals within the quantum scale, which is smaller than a nanometre. It is said that Quantum uncertainty at the microscopic level cannot account for macroscopic nonlinear occurrences, nor can it account for human free will. But, when it comes to the measurement problem, the question of the effect of human free will on the whole universe resurfaces and becomes problematic. The classical concept of time and space along with the classical system of reasoning within the problematic of quantum mechanics has been proven to be inadequate regarding epistemological and ontological manifestations of different interpretations of Quantum physics. Bohr in the so-called Copenhagen interpretation of Quantum Mechanics speculated that ontologically blind chance is at work at the most fundamental level and that the state of a thing is only determined by the act of observation of a human observer. Heisenberg was more cautious and took an ontologically agnostic stance. He believed that we cannot know anything beyond the domain of the observer's observation. In his 1925 milestone paper, he writes: ‘The aim of this work is to set the basis for a theory of quantum mechanics based exclusively on relations between quantities that are in principle observable’ (Rovelli 2018).  Heisenberg expressed the difficulty of shaping our language and thought in agreeing with the observed facts of atomic physics. Regarding the quantum phenomena, detaching subjective and objective aspects of the world proved to be problematic. Our language is unable to describe what happens within atoms; Heisenberg found mathematics free from such limitations. Words or concepts formed by the interplay between the world and us could not be \textit{sharply defined} with respect to their meaning. However, this could only happen where concepts become a part of a system of axioms and definitions expressed consistently by a mathematical scheme (Plotnitsky 2010). Bohr's interpretation and the pragmatic approach of 'shut up and calculate' are still today modus operandi and what is thought to physics students. This approach might try to distance itself from odd and somewhat absurd ideas of Bohr, but it cannot solve the crisis that modern physics brought to our picture of the world, it cannot answer our fundamental questions; whether the world is as it is? And it is as it appears to us? Is such an interpretation of Quantum Mechanics not abandoning physics? \par

\vspace{\baselineskip}On one hand Special Relativity states that no information can travel faster than light, on other hand an important aspect of Quantum Mechanics is Quantum Entanglement in which the state of two particles far away, outside of each other's light cone, i.e. the space-time distance light can travel from a localized single point in space-time to all directions, are entangled in such a way that one determines the other. Einstein called it spooky action at distance. Does the observer's observation truly change the state of events millions of light-years away from us? Einstein was suspicious of Quantum Mechanics, at least he famously objected Bohr's interpretation saying that 'God does not play dice'. He was deeply influenced by Leibniz. Einstein - Podolsky - Rosen paradox, which is a thought experiment, challenged Quantum Mechanics by showing that Quantum Entanglement entails superluminal causality. The paradox concluded that Quantum Mechanics is incomplete and thus there should be some mysterious hidden variables at work. From the beginning of Quantum Mechanics in the late 1920es, there was another interpretation of Quantum Mechanics namely Bohmian Mechanics which also believed that Quantum Mechanics description is incomplete. Bricmont adheres to Bohmian Mechanics and strongly rejects Bohr’s ideas and defends a deterministic interpretation of Quantum Mechanics. We agree with his critique of Bohr, although we find his discussion Realism and Idealism full of over-simplifications and misunderstandings about philosophical views. He himself falls in such rhetoric that is more suited to the Science War rather than an accurate and precise scientific discussion. Bricmont fails to acknowledge that if his rendition of idealism is correct, then why and how a philosopher like Kant has defended Newtonian physics and ‘synthetic a priori’ predicates in natural sciences?\par

\setlength{\parskip}{5.04pt}
\begin{justify}
Bell in his famous theorem proved that the hypothesis of hidden variable contradicts with predications of Quantum Mechanics that are experimentally tested. In 2015, an experimental loophole-free test of Quantum entanglement has been conducted using electron spins separated by 1.3 kilometres (Hensen, Bernien et al. 2015). How can we make sense of Quantum entanglement while superluminal communication is not allowed? There are three proposed hypothesises to answer this dilemma (Myrvold Spring 2019 Edition). The first way is to reject unique outcome and postulate many parallel worlds, according to this hypothesis, the object exists in parallel worlds and the act of observation will locate the observer in one of these worlds, and maybe generate more new worlds. This hypothesis seems to us incompatible with logical principles especially with the principle of contradiction, and cannot explain the identity of objects. The second hypothesis is \textit{retrocausality}, the possibility of causal influence from future to past, which also seems to be incompatible with principle of contradiction and requires the influence of nonbeing over being. The third hypothesis, which Bell himself adheres to and is advocated by Gerard’t Hooft, is \textit{superdeterminism}. From this hypothesis one can conclude that there is no causation at work, as everything is predetermined. This hypothesis is compatible with logical principles. It is only incompatible with our common understanding of our world. This hypothesis is also compatible with Leibniz’s philosophy. It seems to us that modern physics can embrace Leibniz’s philosophy. In 2006, Conway and Kochen (Conway 2006) suggested that electrons have free will. By free will, they do not mean that electrons have consciousness, but rather that electrons do not behave only by mere blind chance, they seem to make a choice in face of choices observer make. If humans have free will, so do electrons. This seems to us the other side of the coin of \textit{superdeterminism}. Conway when explaining their theory in his lectures at Princeton University in 2009, gives an interesting example, he says it is as if watching a movie for the second time. 
\end{justify}\par

\vspace{\baselineskip}
\vspace{\baselineskip}\textbf{References}\par

\vspace{\baselineskip}{\fontsize{11pt}{13.2pt}\selectfont A New System of Nature,$"$  in G. W. Leibniz, \textit{Philosophical Essays}, ed. and trans. R. Ariew and D. Garber (Indianapolis: Hackett, 1989), p. 143.\par}\par

\vspace{\baselineskip}\begin{justify}
{\fontsize{11pt}{13.2pt}\selectfont A.Wheeler, J. (1957). "On the nature of quantum geometrodynamics." \uline{Annals of Physics \textbf{2(6).}}\par}
\end{justify}\par

\begin{justify}
{\fontsize{11pt}{13.2pt}\selectfont Ablowitz, M., Prinari, B (2008). "Nonlinear Schrodinger systems: continuous and discrete." \uline{Scholarpedia.}\par}
\end{justify}\par

\begin{justify}
{\fontsize{11pt}{13.2pt}\selectfont Anishchenko, V. S. (2007). Nonlinear dynamics of chaotic and stochastic systems. Berlin ; New York, Springer\textbf{: xvi, 446 p.}\par}
\end{justify}\par

\begin{justify}
{\fontsize{11pt}{13.2pt}\selectfont Ariew, R. and D. Garber (1989). \uline{G. W. Leibniz Philosophical Essays, Hackett.}\par}
\end{justify}\par

\begin{justify}
{\fontsize{11pt}{13.2pt}\selectfont Bishop, R. (2017). Chaos. \uline{The Stanford Encyclopedia of Philosophy. E. N. Zalta, Metaphysics Research Lab, Stanford University.}\par}
\end{justify}\par

\begin{justify}
{\fontsize{11pt}{13.2pt}\selectfont Bouquiaux, L. (1993). "Monads and Chaos: The Vitality of Leibniz's Philosophy." \uline{Diogenes \textbf{41(161): 87-105.}}\par}
\end{justify}\par

\begin{justify}
{\fontsize{11pt}{13.2pt}\selectfont Bricmont, J. (1995). "Science of Chaos or Chaos in Science?" \uline{ANNALS of the New York Academy of Sciences \textbf{775(1): 131-175.}}\par}
\end{justify}\par

\begin{justify}
{\fontsize{11pt}{13.2pt}\selectfont Callender, C. (Winter 2016 Edition). Thermodynamic Asymmetry in Time. \uline{The\ Stanford Encyclopedia of Philosophy  E. N. Zalta.}\par}
\end{justify}\par

\begin{justify}
{\fontsize{11pt}{13.2pt}\selectfont Conway, J. K., S. (2006). "The Free Will Theorem." \uline{Foundations of Physics \textbf{36(10).}}\par}
\end{justify}\par

\begin{justify}
{\fontsize{11pt}{13.2pt}\selectfont Duncan, S. (2016). Mind and body in early modern philosophy. \uline{Routledge Encyclopedia of Philosophy, Taylor and Francis.}\par}
\end{justify}\par

\begin{justify}
{\fontsize{11pt}{13.2pt}\selectfont Eagle, A. (Spring 2019 Edition). Chance versus Randomness. \uline{The Stanford Encyclopedia of Philosophy. E. N. Zalta.}\par}
\end{justify}\par

\begin{justify}
{\fontsize{11pt}{13.2pt}\selectfont Hájek, A. (Winter 2012 Edition). Interpretations of Probability. \uline{The\ Stanford Encyclopedia of Philosophy  E. N. Zalta.}\par}
\end{justify}\par

\begin{justify}
{\fontsize{11pt}{13.2pt}\selectfont Hensen, B., H. Bernien, A. E. Dréau, A. Reiserer, N. Kalb, M. S. Blok, J. Ruitenberg, R. F. L. Vermeulen, R. N. Schouten, C. Abellán, W. Amaya, V. Pruneri, M. W. Mitchell, M. Markham, D. J. Twitchen, D. Elkouss, S. Wehner, T. H. Taminiau and R. Hanson (2015). "Loophole-free Bell inequality violation using electron spins separated by 1.3 kilometres." \uline{Nature \textbf{526: 682.}}\par}
\end{justify}\par

\begin{justify}
{\fontsize{11pt}{13.2pt}\selectfont Immanuel Kant and T. Paul Guyer (Editor, Allen W. Wood (Editor, Translator) (1999). \uline{Critique of Pure Reason, Cambridge University Press }\par}
\end{justify}\par

\begin{justify}
{\fontsize{11pt}{13.2pt}\selectfont Jaynes, E. T. (1965). "Gibbs vs Boltzmann Entropies." \uline{American Journal of Physics \textbf{33(5): 391-398.}}\par}
\end{justify}\par

\begin{justify}
{\fontsize{11pt}{13.2pt}\selectfont Leibniz, G. W., S. Clarke and R. Ariew (2000). \uline{Correspondence, Hackett Publishing Company.}\par}
\end{justify}\par

\begin{justify}
{\fontsize{11pt}{13.2pt}\selectfont Marqui, L. P. S. (2013). \uline{Philosophical essay on probabilities, Hardpress Limited.}\par}
\end{justify}\par

\begin{justify}
{\fontsize{11pt}{13.2pt}\selectfont Myrvold, W., Genovese, Marco and Shimony, Abner, (Spring 2019 Edition). Bell’s Theorem. \uline{The\ Stanford Encyclopedia of Philosophy  E. N. Zalta.}\par}
\end{justify}\par

\begin{justify}
{\fontsize{11pt}{13.2pt}\selectfont Plotnitsky, A. (2010). \uline{Epistemology and probability : Bohr, Heisenberg, Schrödinger and the nature of quantum-theoretical thinking. New York, Springer.}\par}
\end{justify}\par

\begin{justify}
{\fontsize{11pt}{13.2pt}\selectfont Prigogine, I., I. Stengers and A. Toffler (2018). \uline{Order Out of Chaos: Man's New Dialogue with Nature, Verso Books.}\par}
\end{justify}\par

\begin{justify}
{\fontsize{11pt}{13.2pt}\selectfont Redding,\ P. (Fall 2011). "Journal of Philosophy: A Cross-Disciplinary Inquiry."  \textbf{7(16).}\par}
\end{justify}\par

\begin{justify}
{\fontsize{11pt}{13.2pt}\selectfont Rovelli, C. (2007). \uline{Quantum Gravity, Cambridge University Press.}\par}
\end{justify}\par

\begin{justify}
{\fontsize{11pt}{13.2pt}\selectfont Rovelli, C. (2018). "Physics Needs Philosophy. Philosophy Needs Physics." \uline{Foundations of Physics \textbf{48(5): 481-491.}}\par}
\end{justify}\par

\begin{justify}
{\fontsize{11pt}{13.2pt}\selectfont Sokal, A. D. and J. Bricmont (1998). \uline{Fashionable nonsense : postmodern intellectuals' abuse of science. New York, Picador.}\par}
\end{justify}\par

\begin{justify}
{\fontsize{11pt}{13.2pt}\selectfont Urry, J. (2005). "The Complexity Turn." \uline{Theory Culture Society \textbf{22.}}\par}
\end{justify}\par

\begin{justify}
{\fontsize{11pt}{13.2pt}\selectfont von Leibniz, G. W. F. and R. Latta (1898). \uline{The Monadology and Other Philosophical Writings, Clarendon Press.}\par}
\end{justify}\par

\vspace{\baselineskip}
\printbibliography
\end{document}